\documentstyle[seceq,twocolumn,epsfig]{jpsj}
\newcommand{\ffigureheight}[1]{}
\newcommand{\eepsfig}[1]{\epsfig{#1}}

\def\runtitle{Exact Solution of the Quantum Mattis Model}
\def\runauthor{P. {\sc Sollich}, H. {\sc Nishimori},
A. C. C.  {\sc Coolen}, and A. J. van der {\sc Sijs}}

\newcommand{\bi}{\begin{itemize}}
\newcommand{\ei}{\end{itemize}}
\newcommand{\be}{\begin{equation}}
\newcommand{\ee}{\end{equation}}

\newcommand{\bea}{\begin{eqnarray}}
\newcommand{\eea}{\end{eqnarray}}
\newcommand{\beastar}{\begin{eqnarray*}}
\newcommand{\eeastar}{\end{eqnarray*}}
\newcommand{\lav}{\left\langle}
\newcommand{\rav}{\right\rangle}

\newcommand{\half}{\frac{1}{2}}

\newcommand{\eq}[1]{~(\ref{#1})}

\newcommand{\eqqq}[3]{~(\ref{#1},\ref{#2},\ref{#3})}

\newcommand{\order}{{\cal O}}

\newcommand{\ie}{{\it i.e.}}

\newcommand{\sigv}{\mbox{\boldmath $\sigma$}}
\newcommand{\sv}{\mbox{\boldmath $S$}}
\newcommand{\spv}{\mbox{\boldmath $S$}_+}
\newcommand{\smv}{\mbox{\boldmath $S$}_-}
\newcommand{\bv}{\mbox{\boldmath $B$}}
\newcommand{\db}{\delta\mbox{\boldmath$B$}}

\newcommand{\plus}{\!+\!}
\newcommand{\minus}{\!-\!}

\newcommand{\jm}{\mbox{\boldmath $J$}}
\newcommand{\mmat}{\mbox{\boldmath $M$}}
\newcommand{\hv}{\mbox{\boldmath $h$}}

\newcommand{\mv}{\mbox{\boldmath $m$}}
\newcommand{\mtv}{\mbox{\boldmath $\tilde{m}$}}
\newcommand{\vv}{\mbox{\boldmath $v$}}
\newcommand{\mpv}{\mbox{\boldmath $m$}_+}
\newcommand{\mmv}{\mbox{\boldmath $m$}_-}
\newcommand{\mtmv}{\mbox{\boldmath $\tilde{m}$}_-}
\newcommand{\mpx}{m_+^x}
\newcommand{\mmx}{m_-^x}
\newcommand{\mpy}{m_+^y}
\newcommand{\mmy}{m_-^y}
\newcommand{\mpz}{m_+^z}
\newcommand{\mmz}{m_-^z}
\renewcommand{\mp}{m_+}
\newcommand{\mm}{m_-}
\newcommand{\np}{n_+}
\newcommand{\nm}{n_-}
\newcommand{\Fvar}{\tilde{F}}
\newcommand{\fvar}{\tilde{f}}
\newcommand{\chit}{\mbox{\boldmath $\chi$}}
\newcommand{\zv}{{\rm 0}}
\newcommand{\tensorprod}{\otimes}
\newcommand{\mident}{{\bf 1}}

\newcommand{\Ap}{$\mbox{A}_+$}
\newcommand{\Am}{$\mbox{A}_-$}
\newcommand{\A}{A}
\newcommand{\R}{R}
\newcommand{\vtanh}{\mbox{\boldmath $t$}}

\newcommand{\eps}{\epsilon}
\renewcommand{\d}{\Delta}
\newcommand{\dtri}{\Delta_{\rm tri}}
\newcommand{\dis}{\Delta_{\rm Ising}}
\newcommand{\dthree}{\Delta_3}
\newcommand{\dcep}{\Delta_{\rm CEP}}
\newcommand{\dreentr}{\Delta_{\rm re}}
\newcommand{\btil}{\tilde{B}}
\newcommand{\ttil}{\tilde{T}}

\newcommand{\phip}{\phi_+}
\newcommand{\phim}{\phi_-}

\newcommand{\op}{\psi}
\newcommand{\opv}{\mbox{\boldmath $\psi$}}

\title
{
Exact Solution of the Infinite-Range\\
Quantum Mattis Model
}

\author
{ 
P. {\sc Sollich}, Hidetoshi {\sc Nishimori}$^{1}$,
A. C. C. {\sc Coolen}, and 
A. J. van der {\sc Sijs}$^2$\footnote{Present address:
ASM Litography, PO Box 324, NL-5500 AH
Veldhoven, The Netherlands}
}

\inst
{
Department of Mathematics, King's College London,
Strand, London WC2R 2LS, United Kingdom
\\
$^1$Department of Physics, Tokyo Institute of Technology,
Oh-okayama, Meguro-ku, Tokyo 152-8551
\\
$^2$Department of Physics, University of Oxford, 1 Keble Road, Oxford
OX1 3NP, United Kingdom
}

\recdate
{
\today
}

\abst{
We have solved the quantum version of the Mattis model with
infinite-range interactions. A variational approach gives the exact
solution for the infinite-range system, in spite of the
non-commutative nature of the quantum spin components; this implies
that quantum effects are not predominant in determining the
macroscopic properties of the system. Nevertheless, the model has a
surprisingly rich phase behaviour, exhibiting phase diagrams with
tricritical, three-phase and critical end points.
}

\kword
{
Quantum spin, Mattis model, randomness, infinite-range model,
exact solution, variational method
}

\begin{document}


\sloppy

\maketitle

\section{Introduction}

Quantum spin systems with randomness are of active
current interest because the interplay of
randomness and quantum fluctuations often
leads to nontrivial behaviour
\cite{Bhatt, Nishimori-Nono}.
However, full exact analysis of such systems
is very difficult, partly due to the randomness and
partly due to non-commutativity of
quantum spin operators.
We therefore solve in the present paper a quantum version
of the Mattis model with infinite-range interactions
to investigate the effects of coexistence of randomness
and quantum fluctuations.

The Mattis model was originally proposed as a simple spin glass model
exhibiting only randomness, but no frustration \cite{Mattis}.  It
consists of Ising spins interacting via unfrustrated random exchange
interactions.  The randomness can be gauged away, giving a simple
ferromagnetic Ising model.
The model is nevertheless nontrivial under an external field,
in which case the problem reduces to that of the Ising ferromagnet
with random local fields.
In this paper we study the {\em quantum} version of the Mattis model
in an external field.  This is a more complex problem because, even in
the absence of external fields, one cannot gauge away the randomness
in the exchange interactions without violating the
commutation relations of the quantum spin operators.

There have been several investigations of the quantum Mattis model
including the dispersion relation \cite{sherr} and the 
symmetry of the ground-state.
\cite{nishimori81}
These studies show that the non-commutativity of spin operators
does not necessarily play an important role in the determination
of the qualitative behaviour of the system.
However, there has to date been no explicit solution for the
equilibrium behaviour of the infinite-range model, even without an
external field. The exact solution given in the present paper fills
this gap; it also shows explicitly that, as was to be expected from
the earlier studies cited above \cite{sherr, nishimori81}, quantum
fluctuations do not affect the structure of the phase diagram in an
essential way.
The physical reason for this fact is that the effective field acting
on a spin is the sum of very many other spins when the range of
interactions is infinite; but the sum of many quantum spin operators
behaves like a classical vector.
Hence, as we show, the problem reduces effectively to that of a
single-site quantum spin in a classical external field.

In spite of this fact, the system exhibits very rich phase behaviour.
There are essentially three ordered phases, two with collinear spin
orientations (parallel or antiparallel to the external field), and one
with non-collinear spins.
These phases are separated by first
or second order transition lines which terminate or meet
at critical, tricritical, three-phase or critical end points.
All these types of phase behaviour can be explicitly and exactly
derived by relatively simple but nevertheless nontrivial
manipulations.

The paper is organized as follows.  The model is defined and its
variational free energy derived in \S \ref{sec:model}.  Extremization
of the variational free energy gives the exact solution of the
infinite-range model, even for quantum spins. (In an appendix, we also
confirm the variational result by a direct explicit calculation.) The
three special cases of the Ising, Heisenberg and $XY$ models, which
already exhibit the three types of ordered phases mentioned above, are
treated in \S \ref{sec:special}.
In \S \ref{sec:general}, then, the model is analysed for general
values of the anisotropy parameter in the exchange interaction. While
no additional ordered phases appear, the resulting phase diagrams now
exhibit new and nontrivial features such as tricritical, three-phase
and critical end points. The final section is devoted to conclusions.

\section{Model and free energy}
\label{sec:model}

The system we consider consists of $N$ spin-1/2 quantum spins
$\sv_i$. We use the rescaled spins $\sigv_i=(2/\hbar)\sv_i$,
whose components
have eigenvalues $\sigma_i^\alpha=\pm 1$ ($\alpha=x, y, z$), to
describe the state of the system; this makes correspondences with
classical Ising models easier to see. The Hamiltonian is defined as
\begin{equation}
H = - \frac{1}{N}\sum_{i<j} \xi_i\xi_j\sigv_i\cdot\jm\sigv_j -
\sum_i\bv\cdot \sigv_i.
\label{model}
\end{equation}
Here the sum runs over all pairs of sites, $\jm$ is a general
$3\times 3$ coupling matrix, $\bv$ is an external field, and the
$\xi_i=\pm 1$ are quenched random variables. If only $J_{zz}$ and
$B_z$ are nonzero, then only the $z$-components of the spins,
$\sigma_i^z$, appear in $H$. Because the $\sigma_i^z$ all commute with
each other, the quantum nature of the problem is then irrelevant, and
one recovers the classical Mattis model \cite{Mattis} which is
formulated in terms of Ising spins.

To solve the model\eq{model}, it is tempting to try to gauge away the
quenched disorder by the transformation $\sigv_i \to \xi_i
\sigv_i$. But this is impossible, because the gauged spins would no
longer obey the required 
commutation relations
\be
[\sigma_i^\alpha,\sigma_i^\beta]=2i\sum_\gamma 
\eps_{\alpha\beta\gamma} \sigma_i^\gamma
\label{commutation}
\ee
with $\eps_{\alpha\beta\gamma}$ the fully antisymmetric unit tensor.
Instead, we solve the model using a variational mean field theory
which treats the spins as uncorrelated with each other. Because we are
dealing with a model with infinite range interactions, this
approximation becomes exact in the thermodynamic limit $N\to\infty$;
an explicit self-consistency argument for this fact is given in
Appendix \ref{app:selfconsistency}.
A direct solution of a special case given in Appendix
\ref{app:special_cases} also confirms the variational result.
In the mean field approach, we start
from a trial Hamiltonian
\[
H_0 = - \sum_i \hv_i \cdot \sigv_i
\]
with the associated variational free energy
\[
\Fvar = F_0 + \lav H-H_0 \rav_0.
\]
Here $\lav\ldots\rav_0$ denotes an average over the Boltzmann
distribution defined by $H_0$, and $F_0$ is the corresponding free
energy. The fact that mean field theory is exact then means that the
minimum of $\Fvar$ with respect to the $\hv_i$ is not just an upper
bound on the true free energy $F$, but in fact equal to it in the
thermodynamic limit.  One easily evaluates $\Fvar$ as
\begin{eqnarray}
\Fvar &=& - T \sum_i \ln 2\cosh \beta|\hv_i|
  \nonumber\\
 && -\frac{1}{N}\sum_{i<j} \xi_i\xi_j\mv_i\cdot\jm\mv_j -
\sum_i (\bv-\hv_i)\cdot \mv_i,
\nonumber
\end{eqnarray}
where $\beta=1/T$ as usual.
The magnetizations are given by
\[
\mv_i = \lav\sigv_i\rav_0 = \vtanh(\beta\hv_i)
\]
in terms of the ``vector hyperbolic tangent''
\[
\vtanh(\vv) = \frac{\vv}{|\vv|}\tanh(|\vv|).
\]
Minimizing $\Fvar$ with respect to the $\hv_i$ gives the conditions
\be
\chit_i
\left[\hv_i - \bv - \xi_i\frac{1}{N}\sum_{j\neq i} \xi_j \jm \mv_j \right]
= \zv,
\label{min_first}
\ee
where the $\chit_i$ are the local susceptibility tensors
\be
\chit_i = \frac{\partial \mv_i}{\partial \hv_i}
\label{chit}
\ee
with components given by $\chi_i^{\alpha\beta}$ $=$ ${\partial
m_i^\alpha}/{\partial h_i^\beta}$.  It is easy to show that the
$\chit_i$ are positive definite and hence invertible; explicitly, one
finds ($h_i = |\hv_i|$)
\[
T\chit_i = \frac{\hv_i \tensorprod \hv_i}{h_i^2} (1-\tanh^2 \beta
h_i) + \left(\mident
-\frac{\hv_i \tensorprod \hv_i}{h_i^2}\right) \frac{\tanh \beta
h_i}{\beta h_i}
\]
with $\mident$ denoting the $3\times 3$ identity matrix. Hence the
minimization conditions\eq{min_first} can be written as
\begin{eqnarray}
\hv_i &=& \bv + \xi_i\frac{1}{N}\sum_{j\neq i} \xi_j \jm \mv_j
  \nonumber \\
  &=&
 \bv + \xi_i \jm (\np\mpv - \nm\mmv) - \frac{1}{N}\jm\mv_i,
 \label{mf_eqns}
\end{eqnarray}
where
\[
\mv_\pm = \frac{1}{Nn_\pm} \sum_{i\in I_\pm} \mv_i
\]
are the magnetizations of the sublattices $I_\pm=\{i:\xi_i=\pm 1\}$,
respectively; $\np$ and $\nm$ give the fraction of spins contained in
the two sublattices. The last contribution to $\hv_i$ in\eq{mf_eqns}
can be neglected in the thermodynamic limit, and so the local
variational fields become dependent only on the sublattice, not the
site $i$ itself: all the spins in the same sublattice feel the same
local field and hence have the same magnetization. One can now rewrite
the variational free energy per spin, $\fvar=\Fvar/N$, in terms of the
sublattice magnetizations alone, with the result
\begin{eqnarray}
&&\fvar(\mpv,\mmv)= -T\np s(|\mpv|) -T\nm s(|\mmv|)
 \nonumber\\
& &{}-{} \half
(\np\mpv-\nm\mmv) \cdot \jm (\np\mpv-\nm\mmv) \nonumber\\
& &{}-{} \bv \cdot (\np\mpv+\nm\mmv),
\label{Fvar}
\end{eqnarray}
where the entropic contribution is expressed as usual in terms of the
entropy of a binary distribution,
\be
s(m) = -\frac{1+m}{2}\ln\frac{1+m}{2} - \frac{1-m}{2}\ln\frac{1-m}{2}.
\label{entropy}
\ee
The true free energy per spin is obtained as the minimum of the
variational free energy, hence the final result
\begin{equation}
f = \min_{\mpv,\mmv} \fvar(\mpv,\mmv).
\label{F}
\end{equation}
In the following, because the variational free energy is exact for
$N\to\infty$, we drop the tilde on $\fvar$. The requirement that
$f(\mpv,\mmv)$ must be stationary with respect to $\mpv$ and $\mmv$ gives two
self-consistency equations which can be written in the compact (and
intuitively appealing) form
\bea
\mpv &=& \vtanh[(\np\jm\mpv-\nm\jm\mmv+\bv)/T]
\label{speqs_a}
\\
\mmv &=& \vtanh[(-\np\jm\mpv+\nm\jm\mmv+\bv)/T]
\label{speqs_b}
\eea
%

We emphasize that the energetic terms in the free energy\eq{Fvar} are
exactly of the form that one would expect if the sublattice spins
\[
\sigv_\pm = \frac{1}{Nn_\pm} \sum_{i\in I_\pm} \sigv_i
\]
were classical vectors rather than quantum operators. Working out
commutation relations such as
\begin{eqnarray}
[\sigma_+^\alpha,\sigma_+^\beta] &=& 
\frac{1}{N^2\np^2} \sum_{i,j\in I_+} [\sigma_i^\alpha,\sigma_i^\beta]
\nonumber\\
 &=& 
 \frac{1}{N^2\np^2} \sum_{i\in I_+} 2i \sum_\gamma
\eps_{\alpha\beta\gamma} \sigma_i^\gamma 
 \nonumber\\
 &=& \frac{2i}{N\np} 
\sum_\gamma
\eps_{\alpha\beta\gamma} \sigma_+^\gamma
 \nonumber
\end{eqnarray}
indeed shows that in the thermodynamic limit $N\to\infty$ the vectors
$\sigv_\pm$ become classical, with all components commuting with each
other. Note that even though the $\mv_\pm$ are classical, the quantum
(spin) nature of the problem is still reflected in the functional
form\eq{entropy} of the entropic contribution to the free
energy\eq{Fvar}.

Having derived the free energy of our model for general $\jm$ and
$\bv$, we specialize from now on to the case
\be
\jm = \mbox{diag}(1,1,\Delta), \qquad \bv = (0,0,B).
\label{jbchoice}
\ee
The parameter $\Delta$ here interpolates between three important
limits: For $\Delta=0$ and 1, respectively, we have the $XY$ and
Heisenberg versions of the model, while for $\Delta\to\infty$ the
classical (Ising) Mattis model is recovered. We will analyse these
three cases separately first before studying the richer behaviour
obtained for intermediate values of $\Delta$. Because all phase
diagrams are symmetric under $B\to -B$, we generally restrict
ourselves to $B\geq 0$.

Finally, before proceeding, we note that with the choice\eq{jbchoice},
the model has a trivial rotational symmetry in the $xy$-plane. We
break this symmetry by requiring that $\mpy=0$ and $\mpx\geq
0$. Minimization of\eq{Fvar} with respect to rotations of $\mv_-$
implies that the components of $\mv_+$ and $\mv_-$ in the $xy$-plane
are antiparallel to each other, thus also $\mmy=0$, and $\mmx\leq
0$. So the minimization in\eq{F} only has to be carried out over the
four magnetization components $\mpz$, $\mmz$, $\mpx$, $\mmx$, with the
latter being respectively non-negative and non-positive. We also
assume throughout that the positive sublattice $I_+$ contains more
spins than the negative one, \ie, $\np>\nm$, and take $\nm>0$ in order
to exclude the trivial case of an anisotropic quantum ferromagnet
without disorder. Instead of $\np$ and $\nm$, we will sometimes use
the parameter $\eps$, defined by $n_\pm=(1\pm\eps)/2$; $\eps\to 1$
then corresponds to the disorder-free limit, and our assumptions on
$\np$ and $\nm$ translate into $0<\eps<1$.

\section{Ising, Heisenberg and $XY$ models}
\label{sec:special}

\subsection{Ising limit ($\Delta \to \infty$)}
\label{sec:Ising}

In the limit $\Delta\to\infty$, the free energy\eq{Fvar} is minimized
when the sublattice magnetizations point along the $z$-axis. We thus
only have two nonzero order parameters $\mpz$ and $\mmz$ to consider,
and effectively recover the Mattis model with classical Ising spins.
Using $s(m)=s(-m)$, the free energy simplifies to
\begin{eqnarray}
&&f/\Delta = -\ttil \np s(\mpz) -\ttil\nm s(\mmz) 
  \nonumber\\
  &&{}-{} \half
(\np\mpz-\nm\mmz)^2 - \btil (\np\mpz+\nm\mmz),
\label{Ising_Fvar}
\end{eqnarray}
and the stationarity conditions\eq{speqs_a} and\eq{speqs_b} become
\bea
\mpz &=& \tanh[(\np\mpz-\nm\mmz+\btil)/\ttil]
\label{Ising_speqs_a}
\\
\mmz &=& \tanh[(-\np\mpz+\nm\mmz+\btil)/\ttil].
\label{Ising_speqs_b}
\eea
Here we have introduced the rescaled temperature $\ttil=T/\d$ and
field $\btil=T/\d$ which are the relevant control parameters for
$\d\to\infty$.

Considering first the zero temperature limit $\ttil\to 0$, the only
possible solutions of\eq{Ising_speqs_a} and\eq{Ising_speqs_b} are
$m_{\pm}^z=\pm1$, so we only need to compare the four resulting values of
$f$. For large $\btil$, one finds in this way that $\mpz=\mmz=1$. We
call this phase \Ap, to indicate that both sublattice magnetizations
$\mpv$ and $\mmv$ are {\em aligned} along the direction of the field
$\bv$, with both pointing in the same direction.
At $\btil=\np$, there
is a first order transition to $\mpz=1$, $\mmz=-1$; we denote this new
phase \Am\ because the sublattice magnetizations, while still aligned
with the field, now point in opposite directions. Finally, at
$\btil=0$ we have the conventional first order transition where both
magnetizations change sign, and then the mirror image of the \Ap--\Am
transition occurring at $\btil=-\np$. As pointed out earlier, all
phase diagrams of our model have this symmetry under $B\to -B$, so we
will not mention results for $B<0$ in the following.

\begin{figure}
\begin{center}
\eepsfig{file=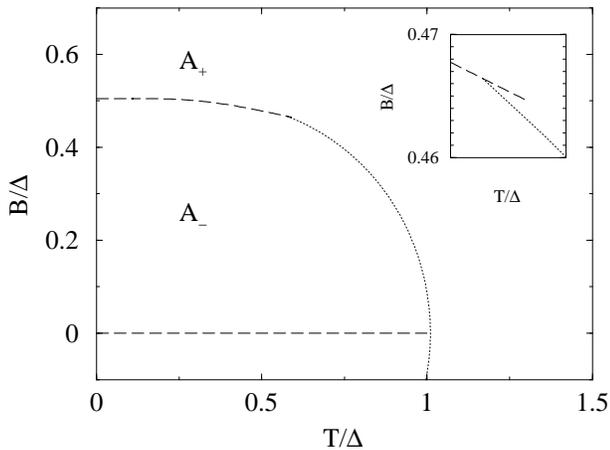,width=8cm}
\end{center}
\ffigureheight{4cm}
\caption{Phase diagram for the Ising limit $\d\to\infty$, for
$\eps=0.01$. The dashed lines indicate first order transitions, and
end in critical points. The transition at $B/\d=0$ and $T/\d<1$ is a
conventional one, where all magnetizations change sign; at the other
transition ($B/\d\neq 0$), the relative orientation of the sublattice
magnetizations changes from antiparallel (\Am) to parallel (\Ap).  The
dotted line indicates the points where the \Am\ and \Ap\ phases
transform smoothly into each other as $\mmz$ passes through zero. As
the inset shows, the \Am--\Ap\ critical point is not on this line; it
has $\mmz>0$.
\label{fig:Ising}
}
\end{figure}

In the $\ttil$-$\btil$ phase diagram, the first order transitions
found above for $\ttil=0$ mark the beginnings of two first order
transition lines, both ending in critical points. The general
procedure by which we find such points is outlined in
Appendix \ref{app:criteria}; after a little algebra, the two relevant
conditions\eq{spinodal_det} and\eq{critical} become
\bea
\ttil &=& \np[1-(\mpz)^2] + \nm[1-(\mmz)^2] \nonumber\\
0     &=& \np\mpz[1-(\mpz)^2] - \nm\mmz[1-(\mmz)^2].
\label{Ising_CP}
\eea
These need to be solved along
with\eq{Ising_speqs_a} and\eq{Ising_speqs_b}. The critical point that marks
the end of the first order transition line at $\btil=0$ has
$\mpz=\mmz=0$ and is thus located at $\ttil=\np+\nm=1$, $\btil=0$. The
critical point terminating the \Ap--\Am\ transition line, on the other
hand, has to be found numerically, along with the location of that
line itself. The resulting phase diagram is shown in
Fig.~\ref{fig:Ising} for $\eps=0.01$.

Note that because of the critical point terminating the line of first
order transitions between \Ap\ and \Am, these two phases can be
smoothly transformed into one another by moving along a continuous
path in the $\ttil$-$\btil$ phase diagram. Along this path, the value
of $\mmz$ changes sign by passing through zero; the points where this
happens obey $\btil=\np\tanh(2\btil/\ttil)$, as is easily derived
from\eq{Ising_speqs_a} and\eq{Ising_speqs_b}. As in the case of the
liquid-gas transition, this makes the thermodynamic distinction
between the two phases somewhat arbitrary. Unlike the traditional
case, however, the critical point is not due to a symmetry under sign
reversal of the magnetization; the critical value of $\mmz$ is
therefore nonzero (and, by\eq{Ising_CP}, positive). This implies that
just below the critical temperature, the \Ap--\Am\ transition is
actually between two phases with $\mmz>0$, rather than two phases with
opposite signs of $\mmz$.

The case of $\epsilon =0$ in the Ising limit is worth an attention
although we generally assume $\epsilon >0$.
When $\epsilon =0$, the system is equivalent to the random-field Ising
model with symmetric distribution of randomness.
This problem has already been solved by the mean-field
approximation.\cite{Aharony}
The phase diagram is similar to Fig. \ref{fig:Ising} except
that the smooth crossover between the \Ap and \Am phases
along the dotted line is now replaced by a second order transition.
This fact can be verified easily, for example, in the case of
$\btil =0$ by setting $n_+=n_-=1/2$ in (\ref{Ising_speqs_a})
and (\ref{Ising_speqs_b}) and rewriting these equations
for the combination $m_+^z-m_-^z$.

Finally, we note that the
results\eqqq{Ising_Fvar}{Ising_speqs_a}{Ising_speqs_b} are valid not
only in the limit $\d\to\infty$, but for {\em all} $\d>0$, as long as
both sublattice magnetizations $\mpv$ and $\mmv$ point along the
$z$-axis. This means that the properties of the \Ap\ and \Am\ phases
are independent of $\d$ when expressed as functions of $\btil$ and
$\ttil$. In particular, the first order \Ap--\Am\ transition as well as
the first order transition at $\btil=0$, $\ttil<1$ will be present in
all phase diagrams unless ``masked'' by other phases.

\subsection{Heisenberg model ($\Delta=1$)}
\label{subsec:Heisengerg}

\begin{figure}
\begin{center}
\eepsfig{file=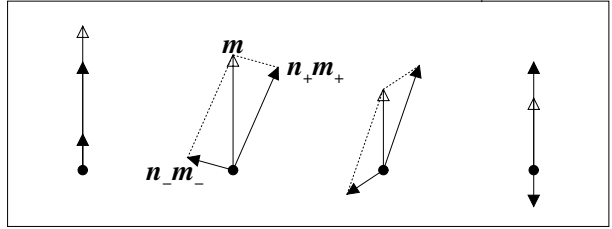,width=8cm}
\end{center}
\ffigureheight{4cm}
\caption{Geometry of the magnetizations for the Heisenberg
model. Shown are (in the $xy$-plane) the total magnetization
$\mv=\np\mpv+\nm\mmv$ and the sublattice magnetizations scaled by the
fractions of spins in the two sublattices, $\np\mpv$ and
$\nm\mmv$. The moduli of the sublattice magnetizations $\mp=|\mpv|$
and $\mm=|\mmv|$ are kept constant while $m=|\mv|$ is decreased from
left to right. For the maximum value of $m$, we have an \Ap\ phase
(left), then we obtain an \R\ phase, with $\mmv$ rotating from the
positive (upward) towards the negative (downward) $z$-axis. Finally,
when $m$ assumes its minimal value (right), we obtain an \Am\
phase. This kind of sequence is observed when passing through the \R\
phase in the Heisenberg phase diagram by decreasing the field $B$ at
constant $T$; the moduli of the sublattice magnetizations remain
constant, and they rotate in such a way as to keep $\mv$ pointing
along the direction of the field (the $z$-direction).
\label{fig:geometry}
}
\end{figure}

In the Heisenberg case $\Delta=1$, the coupling matrix\eq{jbchoice} is
isotropic, and the free energy\eq{Fvar} can be rewritten as
\begin{eqnarray}
 f &=& -T\np s(\mp) -T\nm s(\mm) +\half m^2
  \nonumber\\
  &&-\np^2\mp^2-\nm^2\mm^2 - Bm_z,
\end{eqnarray}
where $m_\pm=|\mv_\pm|$, and $m$ and $m_z$ denote the modulus and
$z$-component, respectively, of the average magnetization
\[
\mv = \np\mpv + \nm\mmv
\]
of all spins. The only dependence on the orientation of the
magnetizations is through the last term, which takes its minimal value
$-Bm$ (remember that we assume $B\geq 0$) when $m_z=m$, \ie, when
$\mv$ is parallel to $\bv$. This gives the free energy
\begin{eqnarray}
f &=&  -T\np s(\mp) -T\nm s(\mm) 
+\half m^2
  \nonumber\\
&&-\np^2\mp^2-\nm^2\mm^2 - Bm
\label{Heisenberg_Fvar}
\end{eqnarray}
which needs to be minimized subject to the constraints
$|\np\mp-\nm\mm|\leq m \leq \np\mp+\nm\mm$. Note that the values of
$\mp$, $\mm$ and $m$ determine the orientations of the magnetizations
uniquely. This is clear geometrically because $\np\mpv$, $\nm\mmv$ and
$\mv$ all lie in the same plane (the $xz$-plane, by our assumption
that $\mpy=\mmy=0$) and form a rigid parallelogram; see
Fig.~\ref{fig:geometry}. The angle of a possible rotation of this
parallelogram about the origin is fixed by the requirement that $\mv$
must be parallel to $\bv$ (\ie, along the positive $z$-axis for
$B>0$). The remaining indeterminacy with respect to a reflection about
the $z$-axis is removed by our assumption that $\mpx\geq 0$.

Focussing now on the minimization of\eq{Heisenberg_Fvar} with respect to $m$,
we note that for $B>\nm\mp+\nm\mm$ this minimum occurs at the maximum
value of $m$, $m=\np\mp+\nm\mm$. Geometrically, this implies that both
$\mp$ and $\mm$ are directed along the positive $z$-axis, so we have
an \Ap\ phase. The free energy becomes
\begin{eqnarray}
 f &=&  -T\np s(\mp) -T\nm s(\mm) -\half(\np\mp-\nm\mm)^2 
  \nonumber\\
   &&-B(\np\mp+\nm\mm),
   \nonumber
\end{eqnarray}
and the sublattice magnetizations therefore obey
\bea
\mp &=& \tanh[(\np\mp-\nm\mm+B)/T] \\
\label{s-mag1}
\mm &=& \tanh[(-\np\mp+\nm\mm+B)/T].
\eea
For small fields obeying $B<\np\mp-\nm\mm$, on the other hand, $f$ is
minimized at the minimum value of $m$, $m=\np\mp-\nm\mm$. With $\mmv$
now pointing along the negative $z$-axis, we have an \Am\ phase with
free energy
\begin{eqnarray}
f &=&  -T\np s(\mp) -T\nm s(\mm) -\half(\np\mp+\nm\mm)^2
  \nonumber\\ & & -B(\np\mp-\nm\mm),
  \nonumber
\end{eqnarray}
and corresponding self-consistency equations
\bea
\mp &=& \tanh[(\np\mp+\nm\mm+B)/T] \\
\mm &=& \tanh[(\np\mp+\nm\mm-B)/T].
\label{s-mag2}
\eea
The above results are identical
to\eq{Ising_Fvar},\eq{Ising_speqs_a} and\eq{Ising_speqs_b}
for the Ising case,
bearing in mind that $\d=1$ and $\mp=\mpz$ here, and that
$\mmz=\pm\mm$ in the \Ap\ and \Am\ phases, respectively. This
conclusion agrees with our general statement in \S ~\ref{sec:Ising}
that properties of the \Ap\ and \Am\ phases are independent of $\d$.

For intermediate values of $B$, finally, the minimum of the free
energy\eq{Heisenberg_Fvar} occurs at a non-extremal value of $m$. In
this regime, we have a new, {\em rotated} (\R) phase where neither
$\mpv$ nor $\mmv$ point along the $z$-axis. Minimization
of\eq{Heisenberg_Fvar} with respect to $m$ gives now $m=B$, and thus
\[
f =  -T\np s(\mp) -T\nm s(\mm) -\np^2\mp^2-\nm^2\mm^2 -\half B^2.
\]
Stationarity with respect to $\mp$ and $\mm$ yields the self-consistency
conditions
\bea
\mp &=& \tanh(2\np\mp/T) \nonumber \\
\mm &=& \tanh(2\nm\mm/T)
\label{R_speqs}
\eea
in which the moduli of the sublattice magnetizations are decoupled and
the field $B$ no longer appears. This may seem surprising at first,
but has a simple explanation: From\eq{speqs_a} and\eq{speqs_b}, the
effective fields that determine the sublattice magnetizations are
(using $\jm=\mident$ for $\d=1$)
\[
\hv_\pm = \pm(\np\mpv-\nm\mmv)+\bv .
\]
But in the \R\ phase, $\mv=\bv$, so that
\[
\hv_\pm = \pm(\np\mpv-\nm\mmv)+(\np\mpv+\nm\mmv) = 2n_\pm\mv_\pm
\]
and the coupling of the sublattices and dependence on $\bv$
disappear, in agreement with\eq{R_speqs}. The geometrical implication
of\eq{R_speqs} is that only the orientations but not the moduli of the
sublattice magnetizations change as the \R\ phase is
traversed. Figure \ref{fig:geometry} shows that as $m$ (and thus $B$)
decreases, $\mmv$ rotates away from the positive $z$-axis and towards
the negative $z$-axis; when it reaches the latter, a transition to an
\Am\ phase occurs. The magnetization of the positive sublattice,
$\mpv$, first rotates away from the $z$-axis and then back towards it;
this follows from the fact that $\mv=\np\mp+\nm\mm$ must keep pointing
along the positive $z$-axis.

\begin{figure}
\begin{center}
\eepsfig{file=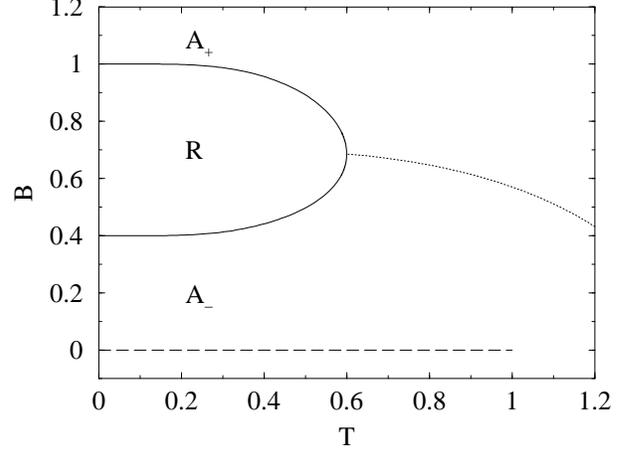,width=8cm}
\end{center}
\ffigureheight{4cm}
\caption{Phase diagram for the Heisenberg case $\d=1$, for
$\eps=0.4$. The dashed line shows the conventional first order
transition at $B=0$, where all magnetizations change sign; the solid
line is a line of second order transitions between aligned (\Ap\ or
\Am) and rotated (\R) phases.
The sublattice magnetization $m_-$ passes smoothly through zero
on the dotted line.
\label{fig:Heisenberg}
}
\end{figure}

Above, the \Ap--\R\ and \R--\Am\ phase boundaries were given as
$B=\np\mp\pm\nm\mm$, respectively. At $T=0$, where $\mp=\mm=1$, these
reduce to $B=\np+\nm=1$ and $B=\np-\nm=\eps$; for nonzero
temperatures, we have to find the phase boundaries numerically by
solving the relevant self-consistency equations for $\mp$ and
$\mm$. We then find (see Fig.~\ref{fig:Heisenberg}) that there is a
line of second order transitions connecting the zero-temperature
\Ap--\R\ and \R--\Am\ transitions; the \R\ phase only occurs inside the
``loop'' formed by this line. The only other feature of the phase
diagram is the first-order transition line at $B=0$, $T<1$, which is
identical to the one found for the Ising case.

\subsection{$XY$ model ($\Delta=0$)}

As the final ``simple'' case we consider the $XY$ version of our
model, which is obtained for $\d=0$. In this case it is convenient to
introduce the mirror image (in the $xy$-plane) of $\mmv$ about the
$z$-axis, given by $\mtmv=(-\mmx,0,\mmz)$, and the corresponding total
(pseudo-) magnetization $\mtv = \np\mpv+\nm\mtmv$. In these variables,
the free energy\eq{Fvar} becomes
\bea
f &=& -T\np s(\mp) -T\nm s(\tilde\mm) 
- \half \tilde m_x^2 - B\tilde m_z \nonumber\\
  &=& -T\np s(\mp) -T\nm s(\tilde\mm)
   \nonumber\\
  && - \half \tilde m^2 + \half \tilde m_z^2 - B\tilde m_z.
\label{XY_Fvar}
\eea
Minimizing over the allowed values of $\tilde m_z\in[-\tilde m,\tilde
m]$ gives $\tilde m_z = B$ as long as $B\leq \tilde m$. The last two
terms in\eq{XY_Fvar} then reduce to the constant $-B^2/2$, and the
remainder of the free energy is minimized (for given $\mp$ and
$\tilde\mm$) when $\tilde m$ takes its maximum value $\tilde m =
\np\mp+\nm\tilde\mm$. Geometrically, this means that $\mpv$,
$\mtmv$ and $\mtv$ are all parallel to each other. The free
energy is
\beastar
f &=& -T\np s(\mp) -T\nm s(\tilde\mm) \\
& & -\half (\np\mp+\nm\tilde\mm)^2 - \half B^2,
\eeastar
while the stationarity conditions
\beastar
\mp &=& \tanh[(\np\mp+\nm\tilde\mm)/T] \\
\tilde\mm &=& \tanh[(\np\mp+\nm\tilde\mm)/T]
\eeastar
show that $\mp=\tilde\mm$, hence also $\tilde m = \mp=\tilde\mm$ with
$\tilde m = \tanh(\tilde m/T)$. The three vectors $\mpv$, $\mtmv$ and
$\mtmv$ are therefore not just parallel, but in fact identical; their
orientation in the $xz$-plane is given by the ratio of $\tilde m_z=B$
and $\tilde m$.  Reverting to the original vectors, we have that
$\mpv$ and $\mmv$ are rotated away from the $z$-axis and are mirror
images of each other under a reflection about this axis; we therefore
have an \R\ phase.  The average magnetization $\mv$ always points
along the positive $z$-axis. Starting from $B=0$, $\mmv$ and $\mpv$
are directed along the negative and positive $x$-axes, respectively
(this follows from $\tilde m^z=0$ and $\mpx\geq 0$). As $B$ is
increased, both sublattice magnetizations rotate towards the $z$-axis
which they reach at the point where $B=\tilde m$.

\begin{figure}
\begin{center}
\eepsfig{file=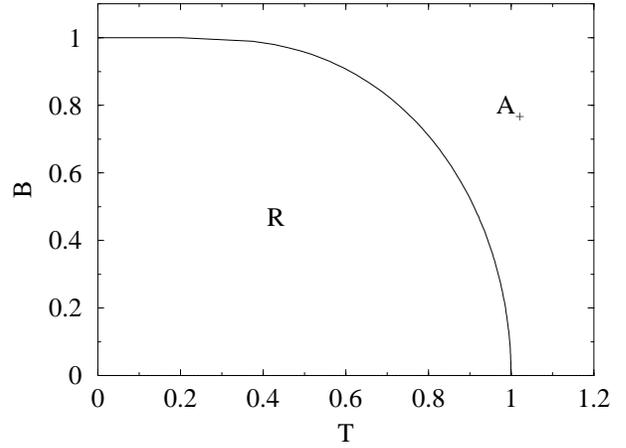,width=8cm}
\end{center}
\ffigureheight{4cm}
\caption{Phase diagram for the $XY$ case $\d=0$.  The solid line
indicates the second order transition between the rotated (\R) and the
aligned (\Ap) phases. Note that this phase diagram is independent of
$\eps$, \ie, of the amount of disorder.
\label{fig:XY}
}
\end{figure}

For larger $B$, we have an \Ap\ phase. In this phase, $\tilde
m_z=\tilde m$; inserting this into\eq{XY_Fvar}, the minimum with respect to
$\tilde m$ is again reached for $\tilde m = \np\mp+\nm\tilde \mm$,
giving
\[
f = -T\np s(\mp) -T\nm s(\tilde\mm) - B(\np\mp+\nm\tilde \mm),
\]
and thus $\mp=\tilde\mm=\tilde m=\tanh(B/T)$. All vectors $\mpv$,
$\mtmv$ and $\mtmv$ are again identical to each other; because they
are now oriented along the $z$-axis, the same is true of the original
magnetizations $\mpv$, $\mmv$ and $\mv$. In Fig.~\ref{fig:XY}, we show
the resulting phase diagram; as expected, there is a line of second
order transitions between the \R\ and \Ap\ phases. Note that obtaining
this line numerically is trivial for the $XY$ model: Combining $\tilde
m =\tanh(\tilde m/T)$ and $B=\tilde m$ gives $B=\tanh(B/T)$. It
follows in particular that in the $XY$ limit, the phase diagram is
actually independent of the amount of disorder (as specified by
$\eps$).

It may appear strange that the disorder has no effect on phase
behaviour in the $XY$ limit $\d=0$. But there is in fact again a very simple
explanation for this. In the model\eq{model}, make a gauge
transformation {\em only on the $x$ and $y$-components} of the spins,
$\sigma_i^x\to\xi_i\sigma_i^x$, $\sigma_i^y\to\xi_i\sigma^y_i$. This
leaves the commutation relations\eq{commutation} unaffected; indeed,
it is just a rotation by $\pi$ around the $z$-axis of the spins with
$\xi=-1$. But this transformation actually gauges away the disorder
completely, so all results must be independent of $\eps$, in agreement
with our findings above. Note that this argument relies on the fact
that only the $x$- and $y$-components of the spins appear in the
disordered (interaction) part of the Hamiltonian\eq{model}, and is
therefore restricted to $\d=0$.

\section{Phase behaviour for general $\Delta$}
\label{sec:general}

We now turn to the study of our model\eq{model} for general values of
$\Delta$. The differences in the phase diagrams for the three cases
studied above ($\Delta=0$, 1 and $\infty$) already suggest that
nontrivial phase behaviour may occur for intermediate values of
$\Delta$. To orient ourselves, we consider first the zero temperature
limit, using $\Delta$ and $B$ as the axes of our phase diagram (and
considering $\eps$ as fixed).

\begin{figure}
\begin{center}
\eepsfig{file=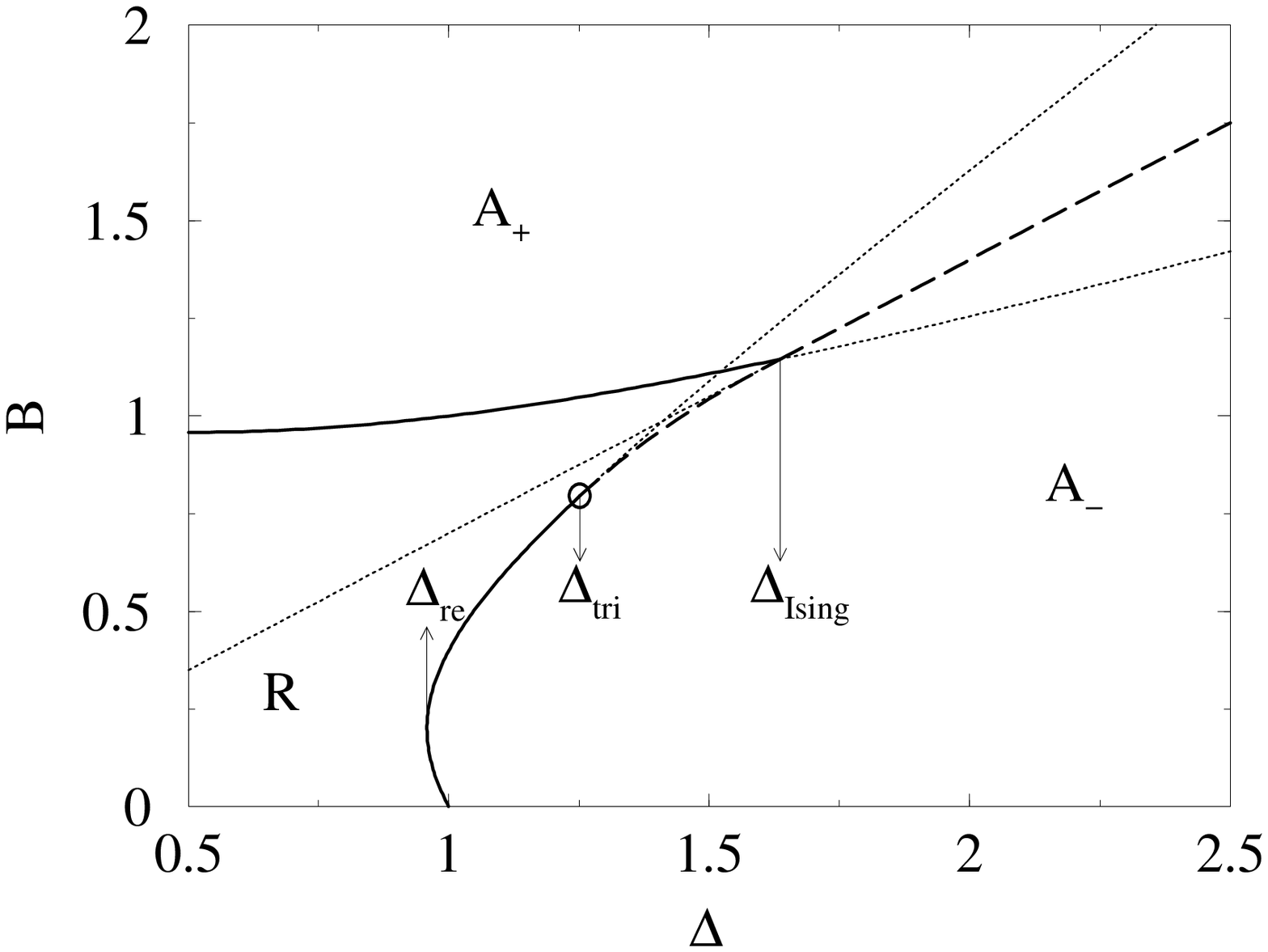,width=8cm}
\end{center}
\ffigureheight{4cm}
\caption{Zero temperature phase diagram for $\eps=0.4$. The
transitions between the three phases \Ap, \Am\ and \R\ are indicated
by bold lines (solid for second order, dashed for first order). Dotted
lines show the continuations of the phase boundaries into metastable
or unstable regimes. Note that the \Am--\R\ transition is re-entrant
for $\dreentr<\d<1$. Also, for $\dtri<\d<\dis$, the first order
\Ap--\Am\ transition is---because of the instability of the \Ap\
phase---pre-empted by a first order transition from \Am\ to \R. Below
$\dtri$, this transition is second order, implying that at $\d=\dtri$
there is a tricritical point (marked by the circle).
\label{fig:zeroT}
}
\end{figure}

For large $\Delta$, we expect essentially Ising behaviour, with an
\Ap\ phase for large $B$, a first order transition to \Am\ at
$B=\np\d$, and the conventional first order transition at $B=0$ where
all magnetizations change sign. In fact, as explained in
\S\ref{sec:Ising}, this will be the zero temperature phase
behaviour for general $\Delta$
unless other phases intervene. We therefore study next the
limits of stability (\ie, the spinodals) of the \Ap\ and \Am\ phases
with respect to a transition to the \R\ phase. This is easiest if we
write the sublattice magnetizations as
\begin{eqnarray}
 \mpv&=&\mp(\sin\phip,0,\cos\phip), \nonumber\\
 \mmv&=&\mm(-\sin\phim,0,\cos\phim).\nonumber
\end{eqnarray}
Here $\phi_\pm$ are the angles (in the $xz$-plane) that $\mpv$ and
$\mmv$ make with the $z$-axis; the signs of the angles were chosen
such that our conventions $\mpx\geq 0$ and $\mmx\leq 0$ always imply
non-negative angles. Using that $\mp=\mm=1$ at $T=0$, the free
energy\eq{Fvar} thus simplifies to
\begin{eqnarray}
 f &=& -\half(\np\sin\phip+\nm\sin\phim)^2
    \nonumber\\
  &&-\half\Delta(\np\cos\phip-\nm\cos\phim)^2 
    \nonumber\\
  &&- B(\np\cos\phip+\nm\cos\phim).
\label{Fvar_zeroT}
\end{eqnarray}
With only two order parameters ($\phip$ and $\phim$) remaining, it is
straightforward to find the matrix of second derivatives of $f$. The
criterion\eq{spinodal_det} implies that a spinodal instability occurs
when the determinant of this matrix vanishes. Evaluating the latter
for $\phip=\phim=0$, one finds for the spinodal of the \Ap\ phase the
condition
\be
B^2-B+\epsilon^2(\Delta-\Delta^2) = 0,
\label{Ap_spinodal}
\ee
while for the \Am\ phase ($\phip=0$, $\phim=\pi$) the corresponding
result is
\be
-B^2+B\epsilon+\Delta^2-\Delta=0.
\label{Am_spinodal}
\ee
The signs of the expressions on the l.h.s.\ have been chosen such that
they are positive when the phases are stable. Also, because of the
symmetry of the free energy\eq{Fvar_zeroT} under $\phip\to-\phip$ and
$\phim\to-\phim$ or $\phim\to2\pi-\phim$, these spinodals
automatically satisfy the critical point criterion\eq{critical} and so
are in fact critical points. Figure \ref{fig:zeroT} shows a plot of the
spinodals lines\eq{Ap_spinodal} and\eq{Am_spinodal} for $\eps=0.4$. We see
that the \Ap\ phase is stable for large fields, but destabilizes as
$B$ is lowered; the \Am\ phase, on the other hand, tends to be stable
for smaller values of $B$, and large $\Delta$. Nontrivially, however,
the \Am\ spinodal shows {\em re-entrance}: For
$\Delta\in[\dreentr,1]$, with
\[
\dreentr = \half(1+\sqrt{1-\epsilon^2}),
\]
Eq.\eq{Am_spinodal} has two physical solutions for $B$, so the \Am\
phase is unstable at zero field, becomes stable at intermediate values
of $B$, and destabilizes again as $B$ is increased further.

In Fig.~\ref{fig:zeroT}, the first order \Ap--\Am\ transition line
$B=\np\d$ is also shown. Moving along this transition line from large
to small $\Delta$, the first spinodal which one crosses is that of the
\Ap\ phase, at the value of $\d$ given by
\[
\dis = 2\,\frac{2\eps+1}{3\eps+1}.
\]
For large values of $\d$, the instabilities of the \Ap\ and \Am\
phases are therefore pre-empted by the first order \Ap--\Am\
transition. In this regime, we expect pure Ising behaviour even at
non-zero temperature, and this is indeed what we find (see below).

Now consider a value of $\d$ just below $\dis$. The \Am\ phase is
stable for $B=0$, and remains so until at $B=\np\d$ the free energy of
the \Ap\ phase becomes lower. The latter is still unstable, however,
because we are below the \Ap\ spinodal. There must therefore be a
stable phase with lower free energy. This phase can only be an \R\
phase (it is neither \Ap\ nor \Am), and a first order transition to
this phase must actually occur at $B<\np\d$. This implies that there
is a line of first order \Am--\R\ transitions extending to the left of
the point $\Delta=\dis,B=\np\dis$. Where this line meets the
\Am\ spinodal (\ie, the line of second order transitions between \Am\
and \R), there will be a {\em tricritical} point. Applying the
criterion\eq{tricritical} to the free energy\eq{Fvar_zeroT}, one finds
after some algebra that this point obeys, in addition
to\eq{Am_spinodal},
\[
B^2 = \d(\d-1)(4\d-3).
\]
It can be shown that, as $\epsilon$ varies between 0 and 1, this
tricritical point moves smoothly from $\Delta=1,B=0$ to
$\Delta=3/2,B=3/2$. In particular, if we call $\dtri$ the value of
$\d$ at the tricritical point, we have $1<\dtri<3/2<\dis<2$ for all
$\eps$.

\begin{figure}
\begin{center}
\eepsfig{file=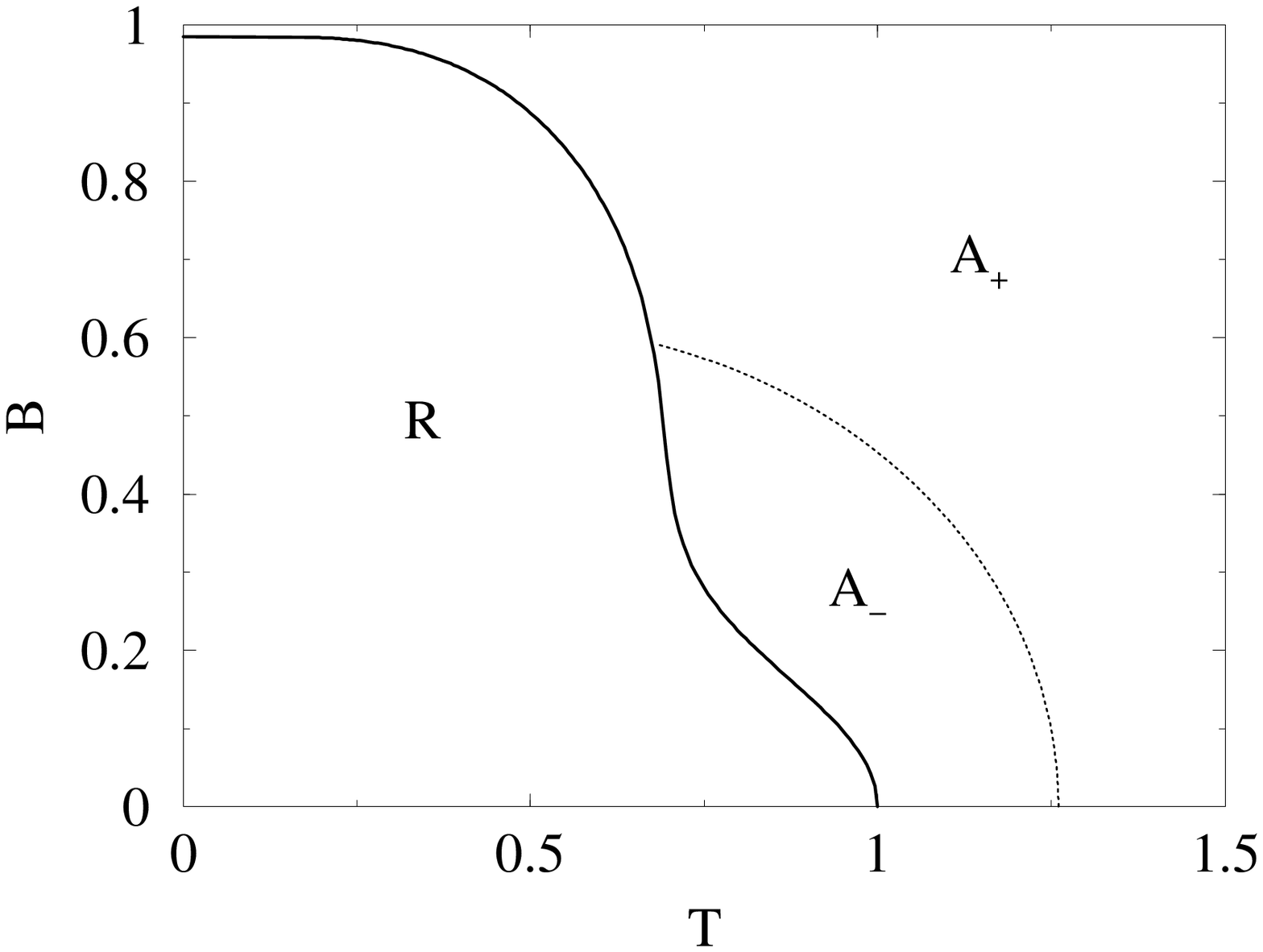,width=8cm}
\eepsfig{file=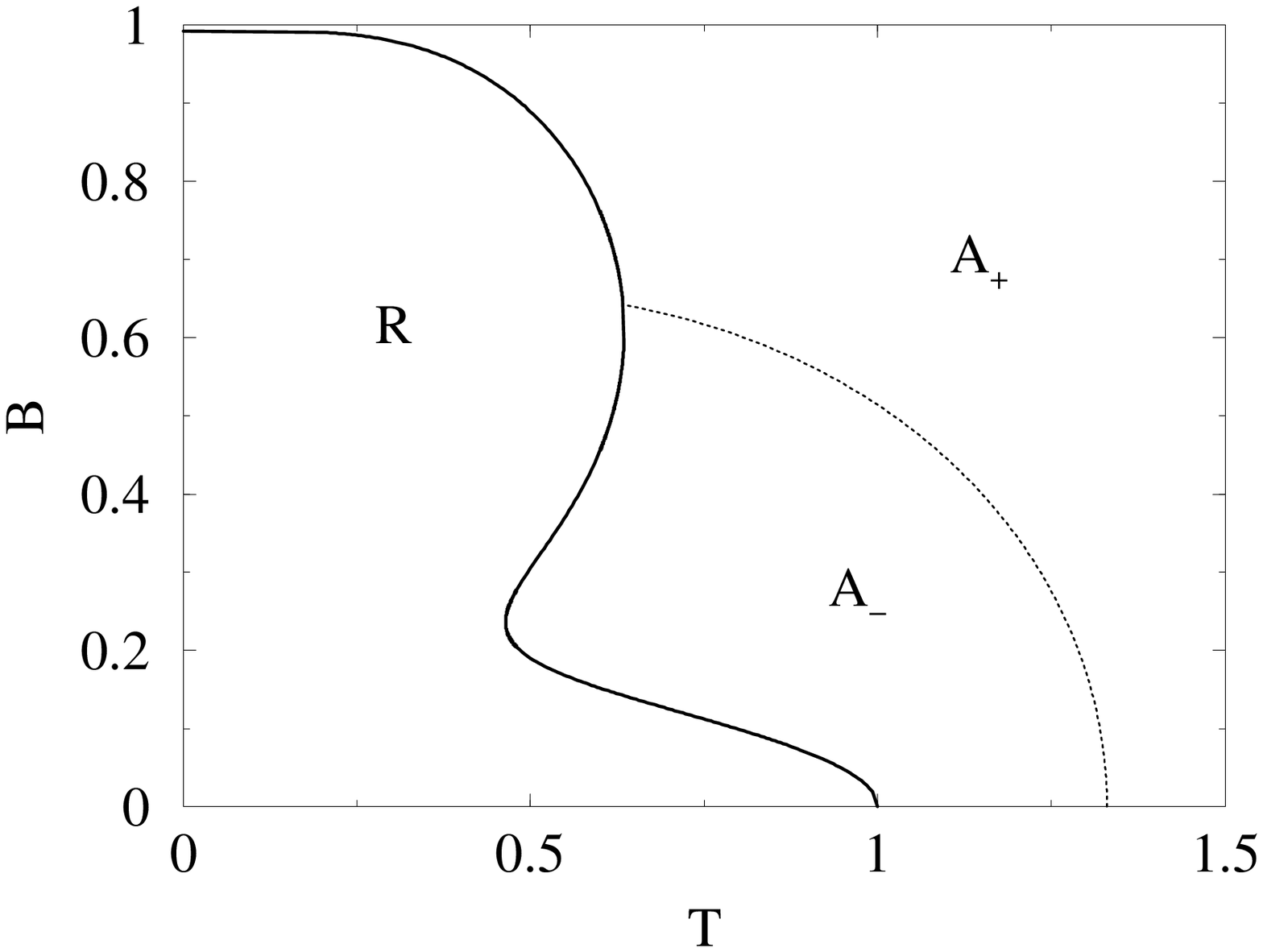,width=8cm}
\end{center}
\ffigureheight{4cm}
\caption{Phase diagrams in regime 1 ($\d<\dreentr$) for
$\eps=0.4$. The solid lines mark second order transitions between the
phases. As in Fig.~\protect\ref{fig:Ising}, the dotted lines show
where the \Am\ and \Ap\ phases transform smoothly into each other
($\mmz=0$). For small $\d$ (top, $\d=0.9$), the phase diagram
resembles qualitatively that of the $XY$ model
(Fig.~\protect\ref{fig:XY}). In the bottom graph, $\d=0.95$ is close
to $\dreentr=0.958\ldots$, and re-entrant behaviour appears at nonzero
temperature. This is a precursor of the transition to regime 2.
\label{fig:XYa}
\label{fig:XYb}
}
\end{figure}

Having clarified the structure of the zero temperature phase diagram,
we can now move on to the finite temperature case. The numerical
results we show were obtained as follows: For spinodals and
tricritical points, we derived analytically the form of the relevant
conditions\eq{spinodal_det} and\eq{tricritical} (for our free
energy\eq{Fvar} with the four order parameters $\mpz$, $\mmz$, $\mpx$
and $\mmx$). We then solved these numerically, along with the
self-consistency equations\eq{speqs_a}. First order transitions were
located as usual by comparing the free energies of the relevant
phases. All results were obtained from double-precision routines for
solving nonlinear simultaneous equations, and cross-checked using a
symbolic manipulation software package with ``arbitrary precision''
floating point operations. We can distinguish a total of seven
different phase diagram topologies, depending on the value of $\d$:

{\em Regime 1: $XY$-like behaviour} ($0\leq \d<\dreentr$).
For small values of $\d$, the phase diagram has essentially the same
features as for the $XY$ limit $\d\to 0$; an example is shown in
Fig.~\ref{fig:XYa}. It is clear that this behaviour cannot persist up
to $\d=1$, however: From the zero temperature phase diagram, we know
that for $\dreentr<\d<1$, there must be re-entrant
behaviour. Correspondingly, the second order \R--\Ap\ transition line
must develop a ``dent''---as if someone was pushing against it from
the positive $T$-direction---as $\d$ increases; this dent will reach
$T=0$ exactly at $\d=\dreentr$. Before this happens, re-entrance will
already be visible for nonzero $T$; Fig.~\ref{fig:XYb} confirms
this. In principle, one could use the appearance of re-entrance at
nonzero $T$ to further divide this regime into two subregimes, but we
choose not to do so because the overall topology of the phase diagram
remains unchanged.

\begin{figure}
\begin{center}
\eepsfig{file=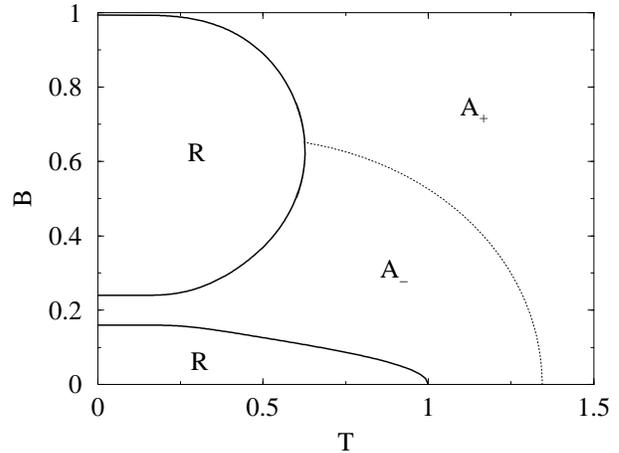,width=8cm}
\end{center}
\ffigureheight{4cm}
\caption{Phase diagram in regime 2 ($\dreentr<\d<1$), for $\eps=0.4$
(where $\dreentr=0.958\ldots$) and $\d=0.96$. The lines have the same
meaning as in Fig.~\ref{fig:XYa}. Re-entrant behaviour now occurs even
at zero temperature.
\label{fig:XY_Heis}
}
\end{figure}

{\em Regime 2: Between $XY$ and Heisenberg} ($\dreentr<\d<1$).
At $\d=\dreentr$, the \R--\Ap\ transition line ``pinches off'' at the
$B$-axis; for larger $\d$, we therefore have two separate transition
lines of this kind (see Fig.~\ref{fig:XY_Heis}). For low enough
temperatures, the sequence of phases observed as $B$ is increased from
0 is therefore \R--\Am--\R--\Ap. As the Heisenberg case $\d=1$ is
approached, the \R--\Am\ transition line (a loop, if we bear in mind
its mirror image for negative $B$) moves closer to the horizontal axis
$B=0$ and at $\d=1$ collapses into the first order transition line of
Fig.~\ref{fig:Heisenberg}. The latter exists for all $\d\geq 1$
(extending up to $\ttil=1$, \ie, $T=\d$, as expected from our
discussion of the Ising case) and will not be mentioned explicitly in
the following.

\begin{figure}
\begin{center}
\eepsfig{file=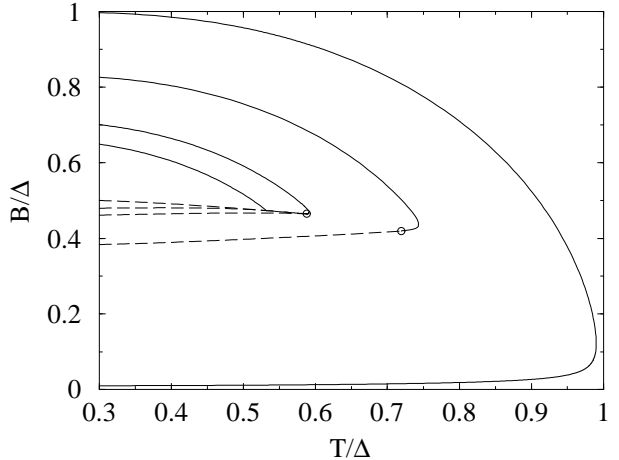,width=8cm}
\end{center}
\ffigureheight{4cm}
\caption{Phase diagrams in regimes 3--7 ($\d=1$, 1.2, 1.4, 1.5, 2 from
the outside to the inside), for $\eps=0.01$. Solid and dashed lines
mark second and first order transitions, respectively; the circle
indicates a tricritical point. See following figures for details.
\label{fig:overall}
}
\end{figure}

{\em Regime 3: Heisenberg-like} ($1\leq\d<\dtri$).
To make the eventual connection with the Ising limit $\d\to\infty$
more apparent, we will use the rescaled variables $\ttil=T/\d$ and
$\btil=B/\d$ as the axes of all phase diagrams from now on. In the
numerical examples, we also switch from $\eps=0.4$ to $\eps=0.01$,
where the intermediate regimes explained below are somewhat easier to
visualize. In the regime $1\leq\d<\dtri$, the phase diagram has
qualitatively the same shape as for the Heisenberg case $\d=1$: A loop
of second order transitions between \R\ and \Ap\ or \Am, respectively,
beginning and ending on the $\btil$-axis (see
Fig.~\ref{fig:overall}). At $\d=\dtri$, the tricritical point that we
found earlier appears on the $\ttil=0$ axis, marking the transition to
the next regime.

{\em Regime 4: Tricritical} ($\dtri<\d<\dthree$).
As $\d$ increases, the tricritical point moves out to larger $\ttil$.
To the left of it, the \Am--\R\ transition is now first order (see
Fig.~\ref{fig:overall}). One might naively expect that as $\d$ is
increased further towards the Ising limit of large $\d$, the ``loop''
of transitions between \R\ and \Am\ or \Ap\ would simply collapse at
$\dis$ onto the Ising first order \Ap--\Am\ transition line. In fact,
two other regimes appear first.

\begin{figure}
\begin{center}
\eepsfig{file=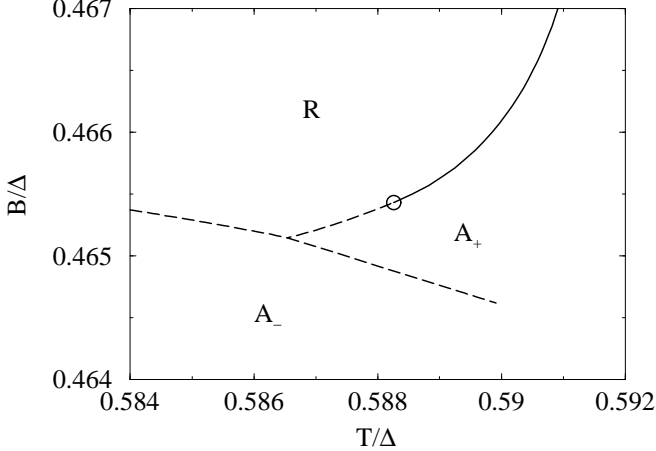,width=8.65cm}
\end{center}
\ffigureheight{4cm}
\caption{Detail of a phase diagram in regime 5 ($\d=1.4$,
$\eps=0.01$). Solid and dashed lines mark second and first order
transitions, respectively; circles indicate tricritical points. There
is a three-phase point where the three first order lines meet.
\label{fig:threephase}
}
\end{figure}

{\em Regime 5: Tricritical plus three-phase} ($\dthree<\d<\dcep$).
It turns out that the \R--\A\ transition loop with the tricritical
point on it does not just shrink along the $\btil$ direction, but also
moves towards smaller values of $\ttil$ as $\d$ is
increased. Eventually, at some $\dthree$, the Ising \Ap--\Am\
transition line therefore ``pokes'' through the loop. This must happen
at a point where the transition is first order; at a point with a
second order transition this would be impossible, as it would imply a
first order transition between two phases which are actually
identical, being both in critical coexistence with the same third
phase. This means that one has a point where three phase transition
lines meet, \ie, a {\em three-phase point}; see
Fig.~\ref{fig:threephase}. There is still a tricritical point where
the \R--\Ap\ phase transition changes from second to first order. Note
that this regime generally corresponds to only a very small range of
$\d$; for $\eps=0.01$, we estimate $\dthree\approx1.39$ and
$\dcep\approx 1.41$.

\begin{figure}
\begin{center}
\eepsfig{file=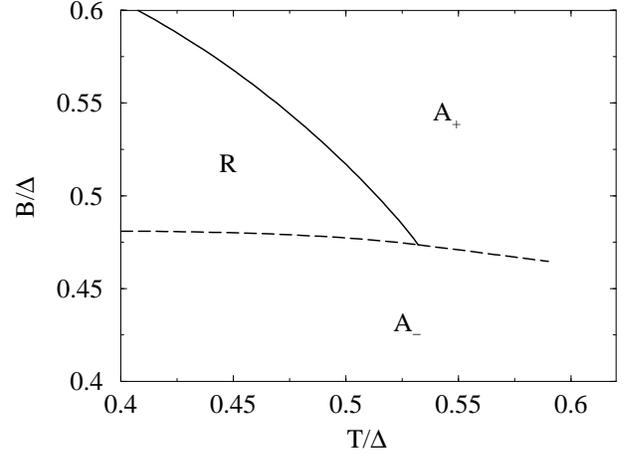,width=8cm}
\end{center}
\ffigureheight{4cm}
\caption{Detail of a phase diagram in regime 6 ($\d=1.5$,
$\eps=0.01$). Solid and dashed lines mark second and first order
transitions, respectively. There is a critical end point where the
line of second order transitions meets the first order line.
\label{fig:CEP}
}
\end{figure}

\begin{figure}
\begin{center}
\eepsfig{file=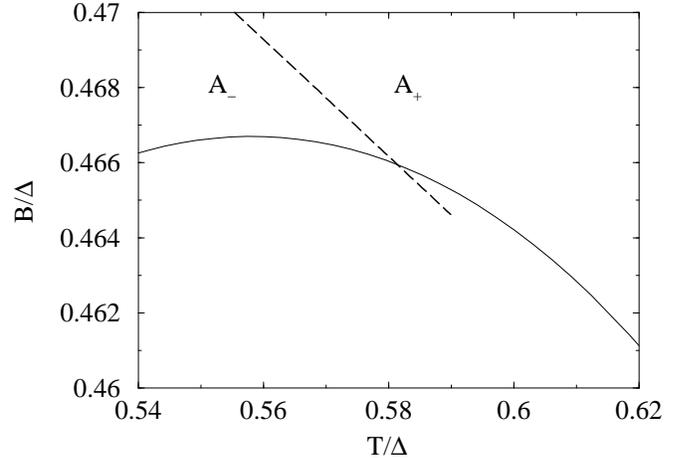,width=8.65cm}
\end{center}
\ffigureheight{4cm}
\caption{The transition between regimes 5 and 6, for $\eps=0.01$. The
dashed line shows the Ising \Ap--\Am\ transition line (which, when
plotted in terms of $\btil=B/\d$ and $\ttil=T/\d$ as done here, is
independent of $\d$). The solid line shows the curve traversed by the
tricritical point as $\d$ is varied, moving from right to left with
increasing $\d$. The crossing of the two curves (\ie, the value of
$\d$ for which the tricritical point meets the \Ap--\Am\ transition and
thus turns into a critical end point) defines $\dcep$. Note that for 
lower values of $\d$ than shown here, the tricritical point would
first continue to move right, but eventually swing back towards the
$\btil$-axis, meeting the latter at $\d=\dtri$.
\label{fig:tricrit_move}
}
\end{figure}

{\em Regime 6: Critical end point} ($\dcep<\d<\dis$).
The relative positions of the tricritical and three-phase points in
the previous regime change as $\d$ increases, until at some $\dcep$
they coincide. From there onwards, the tricritical point is no longer
accessible (it is in a metastable or unstable part of the phase
diagram). Instead, as shown in Fig.~\ref{fig:CEP}, one now has a line
of second order \Ap--\R\ transitions that meets a line of first order
transitions (between \Am\ and \R\ for small $\ttil$, and between \Am\
and \Ap\ for large $\ttil$) at a {\em critical end point}. As $\d$
increases, this point shifts towards lower temperatures and eventually
disappears at $\d=\dis$.
Fig.~\ref{fig:tricrit_move} illustrates the transition between regimes
5 and 6 by showing the curve traversed by the tricritical point as
$\d$ is varied, together with the Ising \Ap--\Am\ transition line. The
value of $\d$ where these two curves meet (\ie, where the tricritical
point ``collides'' with the \Ap--\Am\ transition and thus turns into a
critical end point) defines $\dcep$.

{\em Regime 7: Ising-like} ($\d\geq\dis$).
Finally, for large values of $\d$, one has pure Ising behaviour, with
a $\d$-independent phase diagram (when represented in terms of $\btil$
and $\ttil$) exhibiting the by now familiar line of first order
\Ap--\Am\ transitions (see Fig.~\ref{fig:Ising}).

\section{Conclusion}

We have solved the infinite-range quantum Mattis model
by a variational method that gives the exact solution
in the thermodynamic limit.
The model has various interesting aspects such as
randomness (although without frustration),
quantum effects and competition between
exchange interaction and external field.
We found three ordered phases, two of which have spin states collinear
with the external field and the remaining one with non-collinear
rotated spin states.  The phase diagram has a very rich structure
depending upon the various parameters, in particular the anisotropy of
the interaction.

We now ask how important quantum effects are in producing the
intricate macroscopic behaviour of the system.
The form of the entropy term in the free energy
(\ref{Fvar}) is a direct consequence of the spin-1/2
characteristics of a single quantum spin.
The energy term, on the other hand, is of the form one
would expect classically.
Thus it is clear that the $T=0$ properties do not reflect quantum
effects.  Finite-temperature behaviour, on the other hand, could be
affected by quantum fluctuations.
This fact is somewhat counter-intuitive since quantum effects are
usually most important at {\em low} rather than high temperatures.
An interesting manifestation of finite-temperature quantum effects
is seen in the equations of state (\ref{s-mag1}) to (\ref{s-mag2})
of the Heisenberg
model in the $A_+$ and $A_-$ phases.
These equations of state have exactly the same form as for the Ising
case (\ref{Ising_speqs_a}) and (\ref{Ising_speqs_b}), which would not
be the case if the spins in the Heisenberg model were classical
vectors.
The transverse ({\it i.e.} $x$ and $y$) components of the spin-1/2
operator have disappeared and only the $z$ component comes into play,
as would be expected for a single quantum spin in a classical
field. 

We stress that, even though the quantum nature of the individual spins
in the model does manifest itself at finite temperatures, this does
not produce any qualitatively new ordered phases beyond those already
found at $T=0$. The possible types of ordering are thus determined by
the classical considerations for the ground state; quantum effects
only control how these ordered phases and the boundaries between them
are arranged in the finite temperature phase diagrams.

Finally, a natural question that arises is how our results would
change in the more physically realistic case of finite range
interactions (corresponding to finite dimensionality of the system).
While we cannot give a definite answer to this question at this point,
existing investigations of finite-dimensional quantum Mattis model
\cite{sherr, nishimori81} do suggest that the classical picture gives
reasonable predictions for some features of the model. On the basis of
these results, we conjecture that our observations for the
infinite-range model should give an at least qualitatively reliable
guide to the finite-dimensional problem. Future work is obviously
needed to clarify this point.

\section*{Acknowledgements}
This work was partly supported by the Anglo-Japanese
Scientific Collaboration Programme between the Japan
Society for the Promotion of Science and the Royal
Society.
We thank D. Takigawa for participation in the early stage
of this work.

\appendix

\section{Self-consistency of mean field approximation}
\label{app:selfconsistency}

In this appendix, we show that mean field theory gives exact results
for the generalized Mattis model~(\ref{model}). The intuitive reason
for this is clear: The field that each spin experiences is an average
of $O(N)$ other spins and therefore becomes nonfluctuating (and
classical) in the thermodynamic limit.

To see more explicitly why the mean field approximation, which assumes
that all spins are uncorrelated with each other, becomes
self-consistent for $N\to\infty$, we consider a generalization of our
original Hamiltonian\eq{model} to site-dependent fields:
\[
H = - \frac{1}{N}\sum_{i<j} \xi_i\xi_j\sigv_i\cdot\jm\sigv_j -
\sum_i\bv_i\cdot \sigv_i.
\]
The fluctuation-dissipation theorem then relates the susceptibility
matrix to the spin-spin correlations according to
\be
\frac{\partial \lav\sigv_i\rav}{\partial \bv_j} = -\frac{\partial
F}{\partial \bv_i \partial \bv_j} = \beta \lav
\Delta\sigv_i\Delta\sigv_j \rav, \ \Delta\sigv_i =
\sigv_i-\lav\sigv_i\rav.
\label{FDT}
\ee
For given $i$ and $j$, this is an equality between $3\times 3$
tensors, whose components are written explicitly as $\partial
\lav\sigma_i^\alpha\rav/\partial B_j^\gamma$ $=$ $\beta \lav
\Delta\sigma_i^\alpha \Delta\sigma_j^\gamma \rav$. We can thus verify
that mean field theory is self-consistent by working out the
susceptibility matrix and using it to show that correlations between
different spins vanish for $N\to\infty$.

To obtain the susceptibility matrix, we start from the mean field
equations. By a direct generalization of\eq{mf_eqns}, these are
\be
\hv_i = \bv_i + \xi_i \jm (\np\mpv - \nm\mmv) - \frac{1}{N}\jm\mv_i.
\label{mff_eqns}
\ee
For the case of site-independent fields $\bv_i=\bv$, they reduce
to\eq{mf_eqns}, with the fields $\hv_i$ and magnetizations $\mv_i$
being the same for all spins in each of the two sublattices
$I_\pm$ (in the thermodynamic limit).
We now add a small perturbation $\db$ to the field of one of
the spins; without loss of generality, this spin can be taken as
$\sigv_1$. For definiteness, we also assume that $\sigv_1$ is in the
positive sublattice $I_+$; the calculation for the opposite case is
completely analogous. The solution to the mean field
equations\eq{mff_eqns} will then be such that all spins in the
sublattice $I_-$ still have the same fields $\hv_-$ and magnetizations
$\mv_-$. In $I_+$, on the other hand, we have to distinguish between
the field and magnetization of the chosen spin ($\hv_1$ and $\mv_1$)
and those of all other spins in the sublattice; we denote the latter
by $\hv_2$ and $\mv_2$. Using that the average magnetization of this
sublattice is now
\begin{eqnarray}
\mv_+ &=& \frac{1}{N\np}[\mv_1+(N\np - 1)\mv_2] 
 \nonumber\\
 &=& \frac{1}{N\np}\mv_1 + 
\left(1-\frac{1}{N\np}\right)\mv_2,
\nonumber
\end{eqnarray}
the mean field equations\eq{mff_eqns} then take the form
\begin{full}
\beastar
\hv_1 &=& \bv+\db
+ \jm \left[\frac{1}{N}\mv_1 + \left(\np-\frac{1}{N}\right)\mv_2\right]
- \jm \nm \mv_- -\frac{1}{N}\jm\mv_1 \\
\hv_2 &=& \bv 
+ \jm \left[\frac{1}{N}\mv_1 + \left(\np-\frac{1}{N}\right)\mv_2\right]
- \jm \nm \mv_- -\frac{1}{N}\jm\mv_2 \\
\hv_- &=& \bv 
- \jm \left[\frac{1}{N}\mv_1 + \left(\np-\frac{1}{N}\right)\mv_2\right]
+ \jm \left(\nm - \frac{1}{N}\right) \mv_-.
\eeastar
\end{full}
Subtracting the corresponding equations for $\db=\zv$ gives relations
between the deviations of the fields ($\delta \hv_1$ etc) and
magnetizations ($\delta \mv_1$ etc) from their values for
site-independent fields:
\begin{full}
\beastar
\delta\hv_1 &=& \db
+ \left(\np-\frac{1}{N}\right) \jm\delta\mv_2
- \nm \jm\delta\mv_- \\
\delta\hv_2 &=& \frac{1}{N} \jm\delta\mv_1 +
\left(\np-\frac{2}{N}\right) \jm\delta\mv_2
- \nm \jm \delta\mv_- \\
\delta\hv_- &=& -\frac{1}{N} \jm\delta\mv_1 -
\left(\np-\frac{1}{N}\right) \jm\delta\mv_2
+ \left(\nm - \frac{1}{N}\right) \jm\delta\mv_-.
\eeastar
\end{full}
For small $\db$, we can linearize these using
\[
\delta\mv_1=\chit_+\delta\hv_1, \qquad
\delta\mv_2=\chit_+\delta\hv_2, \qquad
\delta\mv_-=\chit_-\delta\hv_-,
\]
where $\chit_+$ and $\chit_-$ are the local susceptibility
tensors\eq{chit} of spins in the two sublattices; because these are
evaluated for the unperturbed solution (with site-independent fields
$\bv$), there is no need to distinguish between $\chit_1$ and
$\chit_2$. We thus obtain
\begin{full}
\beastar
\chit_+^{-1}\delta\mv_1 - \left(\np-\frac{1}{N}\right)\jm\delta\mv_2 +
\nm \jm\delta\mv_- &=& \db \\
-\frac{1}{N}\jm\delta\mv_1 +
\left[\chit_+^{-1}-\left(\np-\frac{2}{N}\right)\jm\right]\delta\mv_2 +
\nm \jm\delta\mv_- &=& \zv \\
\frac{1}{N}\jm\delta\mv_1 +
\left(\np-\frac{1}{N}\right)\jm\delta\mv_2 +
\left[\chit_-^{-1}-\left(\nm -\frac{1}{N}\right)\jm\right]
\delta\mv_- &=& \zv.
\eeastar
\end{full}
These equations can solved explicitly for the changes in the
magnetizations; keeping only the leading order terms for $N\to\infty$,
one finds
\beastar
\delta\mv_1 &=& \chit_+\db \\
\delta\mv_2 &=&  \frac{1}{N}\chit_+\jm
\left[\mident-\np\chit_+\jm - \nm\chit_-\jm\right]^{-1}\chit_+\db\\
\delta\mv_- &=& - \frac{1}{N}\chit_-\jm
\left[\mident-\np\chit_+\jm - \nm\chit_-\jm\right]^{-1}\chit_+\db.
\eeastar
The $3\times 3$ tensors multiplying $\db$ on the r.h.s.\ give the
entries of the susceptibility matrix $\partial\mv_i/\partial \bv_j$
(for the case $j=1\in I_+$ considered here; as stated above, the case
$j\in I_-$ can be treated in exactly the same fashion). We read off
that these entries are of $\order(1)$ only for $i=j$, while all
off-diagonal terms are $\order(1/N)$. The fluctuation-dissipation
theorem\eq{FDT} then implies that all correlations
$\lav\Delta\sigv_i\Delta\sigv_j\rav$ between different spins $i\neq j$
are $\order(1/N)$; in the thermodynamic limit, mean field theory
therefore becomes exact.

\section{Direct solution for a special case}
\label{app:special_cases}

It is possible in some cases to derive the variational free energy
(\ref{Fvar}) directly from the Hamiltonian (\ref{model}).  We consider
the example of the Heisenberg model ($\Delta =1$) here to confirm the
variational calculations.

Setting $\hbar=1$,
the Hamiltonian (\ref{model}) can be written in terms of total spin
operators as
 \[
  H =
  -\frac{4}{N}\left( \spv^2+\smv^2-\half\sv^2 \right)
           - 2BS_{z},
\]
where we have ignored a trivial constant term.
The spin operators are defined by
 \[
   \sv_{\pm}=\half \sum_{i\in I_\pm} \sigv_i,~~~\sv=\spv+\smv.
 \]
Because the quartet $\{\spv^2,\smv^2,\sv^2,S_z\}$ form
a set of mutually commuting operators,
the Hilbert space of the $N$-particle system
is spanned by their simultaneous eigenspaces $|s_{+},s_{-};s,m\rangle$.
The free energy per spin $f_N$ is therefore
  \begin{eqnarray}
   f_{N}&=&
       -\frac{T}{N}\ln
   \sum_{s_{+},s_{-}}d^{s_{+}}_{N_{+}}d^{s_{-}}_{N_{-}}
   e^{\frac{4\beta}{N}\left[s_{+}(s_{+}+1)+s_{-}(s_{-}+1)\right]}
   \nonumber\\
   &\times &
   \sum_{s=|s_{+}-s_{-}|}^{s_{+}+s_{-}}
   e^{-\frac{2\beta}{N}s(s+1)}\sum_{m=-s}^{s}e^{2m\beta B}
   \label{exact_fren}
  \end{eqnarray}
Here the sums over $s_{\pm}$ run from 0 (or 1/2 if $N_{\pm}$ is odd)
to $N_{\pm}/2$ $(=n_{\pm}N/2)$. The symbol $d^{n}_{M}$ denotes the
number of multiplets of total spin $n$ in a system of $M$
spin-$\frac{1}{2}$ particles. We show in Appendix
\ref{app:combinatorial} that $d^n_M$ is given explicitly by
 \begin{eqnarray}
 d^{n}_{M}&=&
{M-1 \choose \frac{1}{2}M+n-1} -
{M-1 \choose \frac{1}{2}M+n+1}
\nonumber\\
&=& \frac{(2n+1)M!}
{ (\frac{1}{2}M-n)! (\frac{1}{2}M+n+1)! }\ .
 \label{dMn}
 \end{eqnarray}

To evaluate the free energy\eq{exact_fren}, we first carry out the sum
over $m$,
 \[
    \sum_{m=-s}^{s}e^{2m\beta B}=
   \frac{e^{(2s\plus 1)\beta B} - e^{-(2s\plus 1)\beta B}}
{e^{\beta B} - e^{-\beta B}}\ .
 \]
Setting $L=\pm\beta B$, we therefore need to evaluate the following
sum over $s$:
 \begin{eqnarray}
   && \sum_{s=|s_+-s_-|}^{s_{+}+s_{-}}
        e^{-\frac{2\beta}{N}s(s+1)+(2s+1)L}
            \nonumber\\
           && =
    e^L\int\!\frac{dz}{\sqrt{2\pi}}\ e^{-\half z^2}
 \sum_{s=|s_+-s_-|}^{s_{+}+s_{-}}
           e^{2s(L-\beta /N-iz\sqrt{\beta /N})}
         \nonumber\\
  && =e^L
    \int\! \frac{dz}{\sqrt{2\pi}}\ e^{-\half z^2}\times
    \nonumber\\
    &&
    \left[\frac{
       e^{2|s_+ - s_-|(L-\beta /N-iz\sqrt{\beta /N})}-
   e^{2(s_+ + s_- +1)(L-\beta /N-iz\sqrt{\beta /N})}}
      {1-e^{2(L-\beta /N-iz\sqrt{\beta /N})}}\right]
      \nonumber
 \end{eqnarray}
Rescaling the integration variable by $\sqrt{\beta N}$, we thus have
 \begin{eqnarray}
 && \sum_{s=|s_+-s_-|}^{s_{+}+s_{-}}
    e^{-\frac{2\beta }{N}s(s+1)}
   \left[e^{(2s\plus 1)\beta B} - e^{-(2s\plus 1)\beta B}\right]
    \nonumber\\
  &&  =
   \left(\frac{\beta N}{2\pi}\right)^{\frac{1}{2}}
     \int\!dz~e^{-\frac{1}{2}\beta Nz^2}
     \nonumber\\
  && \times
     \left\{
    \frac{e^{2\beta|s_+ - s_-|(B-1/N-iz)}
     -e^{2\beta(s_+ + s_-+1)(B-1/N-iz)}}
    {e^{-\beta B}-e^{\beta(B-2/N-2iz)}}
    \right.\nonumber\\
   && -\left.
    \frac{e^{2\beta|s_+ - s_-|(-B-1/N-iz)}
     -e^{2\beta(s_+ + s_-+1)(-B-1/N-iz)}}
         {e^{\beta B}-e^{\beta(-B-2/N-2iz)}}
   \right\}
   .
   \nonumber
\end{eqnarray}
The thermodynamic limit of the free energy is then expressed as
 \begin{eqnarray}
   && f=- \lim_{N\to\infty}
      \frac{T}{N}\ln
   \int\!dz~e^{-\frac{1}{2}\beta Nz^2}
   \nonumber\\
   &&\times
    \sum_{s_{+},s_{-}}d^{s_{+}}_{N_{+}}d^{s_{-}}_{N_{-}}
    e^{\frac{4\beta}{N}\left[s_{+}(s_{+}+1)+s_{-}(s_{-}+1)\right]}
   \nonumber\\
  &&  \times
  \left\{
       \frac{e^{2\beta |s_+ - s_-|(B-iz)}-e^{2\beta(s_+ +
           s_-+1)(B-iz)}}
         {e^{-\beta B}-e^{\beta (B-2iz)}}
         \right. \nonumber\\
         && \left.
         {}-
        \frac{e^{2\beta|s_+ - s_-|(-B-iz)}-e^{2\beta(s_+ +
           s_-+1)(-B-iz)}}
         {e^{\beta B}-e^{\beta(-B-2iz)}}
    \right\}.
    \nonumber
\end{eqnarray}
The sum over $s_{\pm}$ can be replaced by an integral
in the thermodynamic limit:
 \[
   s_{\pm}=\frac{1}{2}n_{\pm} m_\pm N,~~~~~~
   \sum_{s_{\pm}}\rightarrow\frac{1}{2}n_{\pm} N \int_0^1\! dm_\pm .
 \]
Also, for large $N$ the expression\eq{dMn} for the combinatorial terms
$d^{n}_{N}$ can be simplified to
 \begin{eqnarray}
  \frac{1}{N}\ln d^{s_{\pm}}_{N_{\pm}}
    &=&
  \frac{1}{N}\ln \left\{ 
        \frac{N_\pm !}
      {[N_\pm(\frac{1}{2}\minus m_\pm)]! [N_\pm(\frac{1}{2}\plus
        m_\pm)]!}
	\times \right.
          \nonumber\\
& & \left.
\frac{2N_\pm m_\pm\plus 1}{N_\pm(\frac{1}{2}\plus m_\pm)\plus 1}
\right\}
          \nonumber\\
  &=&
  n_{\pm}s(m_\pm) + {\mathcal O}(1/N).
          \nonumber
  \end{eqnarray}
The free energy thus becomes
  \[
    f=- \lim_{N\to\infty}
  \frac{T}{N}\ln
  \int_{-\infty}^\infty\!dz\int_0^1\!dm_+\int_0^1\!dm_-~
   e^{-\beta N f(z,m_+,m_-)}
  \]
with
\begin{full}
  \begin{eqnarray}
  &&f(z,m_+,m_-)=\frac{1}{2} z^2
  -n_{+}^2 m_{+}^2-n_{-}^2 m_{-}^2-T n_{+} s(m_+)-T n_{-} s(m_-)
   \nonumber\\
&&
  -\lim_{N\to\infty}\frac{T}{N}\ln
  \left\{
  \frac{e^{\beta N|n_{+}m_{+} -n_{-}m_{-}| (-iz+B)}}
          {\sinh[\beta(iz\minus B)]}
 -\frac{e^{\beta N(n_{+}m_{+} +n_{-}m_{-}) (-iz+B)}}
          {\sinh[\beta(iz\minus B)]}
   \right.
         \nonumber\\
&&
   \left.
   -\frac{e^{\beta N|n_{+}m_{+} -n_{-}m_{-}| (-iz-B)}}
          {\sinh[\beta(iz\plus B)]}
   +\frac{e^{\beta N(n_{+}m_{+} +n_{-}m_{-}) (-iz-B)}}
          {\sinh[\beta(iz\plus B)]}
     \right\}.
         \nonumber
\end{eqnarray}
\end{full}
For $N\to\infty$, the third and fourth terms in the curly brackets
become exponentially small compared to the first and second terms,
respectively, and can therefore be discarded (remember that we assume
$B>0$). If we also make the change of variable $z\to iz$ (which
implies that the free energy is to be maximized with respect to $z$),
we have
\begin{full}
  \begin{eqnarray}
  &&f(z,m_+,m_-)=-\frac{1}{2} z^2
  -n_{+}^2 m_{+}^2-n_{-}^2 m_{-}^2-T n_{+} s(m_+)-T n_{-} s(m_-)
   \nonumber\\
&&
  -\lim_{K\to\infty}\frac{1}{K}\ln
  \left\{
\frac{ e^{K|n_{+}m_{+} -n_{-}m_{-}| (z+B)} }
          {\sinh[-\beta(z\plus B)]}
+ \frac{ e^{K(n_{+}m_{+} +n_{-}m_{-}) (z+B)} }
          {\sinh[\beta(z\plus B)]}
   \right\}.
         \nonumber
\end{eqnarray}
\end{full}
Because $n_{+}m_{+} +n_{-}m_{-}\geq |n_{+}m_{+} -n_{-}m_{-}|$, the
first term inside the curly brackets becomes negligible for $z>-B$;
conversely, for $z<-B$, the second term can be discarded. We are then
left with
 \begin{eqnarray}
  &&f(z,m_+,m_-)=
   -\frac{1}{2} z^2
  -n_{+}^2 m_{+}^2-n_{-}^2 m_{-}^2
  \nonumber\\
  &&-T n_{+} s(m_+)-T n_{-} s(m_-)
   \nonumber\\
  &&- (z+B)\times\left\{
    \begin{array}{lll}
           n_{+}m_{+}+n_{-}m_{-}
                  && (z>-B)\\
           |n_{+}m_{+}-n_{-}m_{-}|
                  && (z<-B)
    \end{array}
    \right.
    .
    \label{fr-1}
  \end{eqnarray}
Taking the derivative of $f$ with respect to $z$ subsequently
gives the equations
$z=-(n_{+}m_{+}+n_{-}m_{-})$ if $z>-B$ and
$z=-|n_{+}m_{+}-n_{-}m_{-}|$ if $z<-B$.
In addition we have a local maximum at $z=-B$ if the
$z$-derivative of $f$ is negative for $z>-B$ and positive for
$z<-B$, which translates into the extra solution $z=-B$
appearing for
$|n_{+}m_{+}-n_{-}m_{-}|<B<n_{+}m_{+}+n_{-}m_{-}$.
In combination the full picture now becomes:
\begin{full}
\begin{equation}
  \begin{array}{lll}
   B<|n_{+}m_{+}-n_{-}m_{-}|:&&
       z=-|n_{+}m_{+}-n_{-}m_{-}|\\
   |n_{+}m_{+}-n_{-}m_{-}|< B <n_{+}m_{+}+n_{-}m_{-}:&&
       z=-B\\
   B>n_{+}m_{+}+n_{-}m_{-}:&&
       z=-(n_{+}m_{+}+n_{-}m_{-})
  \end{array}
  .
  \label{regions}
\end{equation}
\end{full}
Elimination of
the variable $z$ using the above result leads us to a reduced free energy
minimization problem involving $m_\pm$ only, with
\begin{full}
\begin{equation}
  \begin{array}{lll}
   (a) & {\rm Phase~A}_{-} : & B<|n_{+}m_{+}-n_{-}m_{-}|\\
   (b) & {\rm Phase~R} : & |n_{+}m_{+}-n_{-}m_{-}|< B <
                          n_{+}m_{+}+n_{-}m_{-}\\
   (c) &  {\rm \rm Phase~A}_{+} : & B > n_{+}m_{+}+n_{-}m_{-}
  \end{array}
  \label{regions2}
\end{equation}
\end{full}
It is straightforward to verify from Eqs. (\ref{fr-1}),
(\ref{regions}) and (\ref{regions2}) that the free energy in each
phase agrees with that given in \S \ref{subsec:Heisengerg}.

\section{Degeneracy factor $d_{M}^{n}$}
\label{app:combinatorial}

We derive here the expression\eq{dMn} for the number of multiplets of
total spin $n$ in a system of $M$ spin-$\frac{1}{2}$ particles.  In
other words, $d_M^n$ is the degeneracy of the state $|n,n_z\rangle$,
where $n$ is the total spin quantum number and $n_z$ is any of the
allowed values ($-n$, $-n+1$, \ldots, $n$) of the $z$-component of the
total spin.

We proceed by induction over $M$. For $M=1$, we have $d^n_M=1$ for
$n=1/2$ and $d^n_M=0$ otherwise. Now consider a multiplet of spin $n$
in an $M$-spin system.  When one spin-1/2 particle is added, this
multiplet splits into exactly one $n+\half$ and one $n-\half$
multiplet.
The exception is $n=0$, where only a single $n=\half$
multiplet is generated.
We therefore have
 \begin{equation}
   d_{M+1}^{n}=d_{M}^{n-\half}+d_{M}^{n+\half}
   \label{recursion}
 \end{equation}
with the boundary condition
 \begin{equation}
   d_{M}^{n}=0~~~(n\le -\mbox{$\textstyle\half$} ).
   \label{r-boundary}
 \end{equation}
Now consider an unbiased random walk with sites
numbered by the integers $l=2n+1$ and discrete time $t=M-1$.
The recursion (\ref{recursion}) then tells us that
 \begin{equation}
   2^{-(M-1)}d_{M}^{n}=p_t(l),
  \label{dprelation}
 \end{equation}
where $p_t(l)$ is the site occupation probability at time $t$ of
a random walk starting from initial position $l_0=2$ at $t=0$.
The boundary condition (\ref{r-boundary}) simply
corresponds to an absorbing wall at $l=0$.
Without this absorbing wall, one would have
  \[
    p_t(l)=2^{-t}
{ t \choose \half (l-l_0+t)}\ ;
   \]
in the presence of
the wall, one simply subtracts the reflected solution in the usual way
to get
  \[
    p_t(l)=2^{-t} \left[ 
{ t \choose \half (l-l_0+t) } -
{ t \choose \half (l+l_0+t) }
     \right].
   \]
Inserting this into (\ref{dprelation}) immediately
gives the desired result\eq{dMn}. In writing the above formulae, we
use the convention that the binomial coefficient ${n \choose k}$ is
zero whenever $k$ is non-integer or outside the range $0\ldots n$.

\section{Criteria for spinodals, critical and tricritical points}
\label{app:criteria}

In this appendix, we set out the general criteria that we use to find
spinodals, critical points and tricritical points. Rather than the
traditional determinant conditions due to Gibbs~\cite{Gibbs}, we use a
formulation due to Brannock~\cite{Brannock} which is more convenient,
especially for tricritical points.

Let us assume the free energy per spin is given by $f(\opv)$, where
$\opv=(\op_1\ldots \op_n)$ is a vector of $n$ (non-conserved) order
parameters. Thermodynamic phases correspond to (local) minima of
$f(\opv)$ and therefore obey the stationary condition $\nabla
f(\opv)=\zv$. At a spinodal point, there is in addition an instability
direction $\delta\opv$ along which the free energy has zero curvature,
implying that the gradient of $f$ remains zero to first order:
\be
(\delta\opv\cdot\nabla)\nabla f(\opv)=\zv \qquad \mbox{(spinodal)}.
\label{spinodal}
\ee
The condition for such a $\delta\opv$ to exist is 
\be
|\mmat|=0, \qquad \mmat = \nabla\nabla f(\opv).
\label{spinodal_det}
\ee
At a critical point, the separation between two neighbouring (in the
$\opv$-space) stable phases becomes infinitesimal. These phases are
separated by an unstable phase (a saddle point of $f$). Constructing a
curve $\opv(s)$ through these three phases (with $\opv(0)=\opv$, the
state we are interested in), we see that (the vector-valued function)
$\nabla f(\opv(s))$ vanishes at three infinitesimally separated values
of $s$. At the critical point, located at $s=0$, these three zeros of
$\nabla f(\opv(s))$ coincide, so
\[
\nabla f(\opv(s)) = \order(s^3).
\]
Similarly, at a tricritical point we have three stable phases coming
together, with two unstable phases between them, so the corresponding
criterion is
\[
\nabla f(\opv(s)) = \order(s^5).
\]
Noting that the spinodal criterion\eq{spinodal} can be written as
$\nabla f(\opv(s)) = \order(s^2)$ for the curve
$\opv(s)=\opv+s\delta\opv$, we can summarize all three criteria as
follows: If there exists a curve $\opv(s)$ through the point $\opv$
(with $\opv(0)=\opv$) such that
\be
\nabla f(\opv(s)) = \order(s^l),
\label{general_crit}
\ee
then for $l=2$, 3, 5 respectively $\opv$ is a spinodal, critical and
tricritical point. 

To evaluate the criterion\eq{general_crit} in practice, we write it as
\be
\left. 
\left(\frac{d}{ds}\right)^k\nabla f(\opv(s)) \right|_{s=0}
= \zv \qquad \mbox{for } k=1\ldots l-1.
\label{deriv_k}
\ee
For the spinodal criterion, this reduces to\eq{spinodal} if we
identify $\delta\opv$ and $\opv'(0)$. For critical points ($l=3$) one
obtains the additional equation
\[
\nabla(\delta\opv\cdot\nabla)^2 f(\opv) +
\nabla(\opv''(0)\cdot\nabla) f(\opv) = \zv.
\]
Taking the scalar product with $\delta\opv$ and using\eq{spinodal}
eliminates the second term, showing that the criterion for critical
points is
\be
(\delta\opv\cdot\nabla)^3 f(\opv) = \zv
\label{critical}
\ee
together with\eq{spinodal}. Following a similar procedure, one finds
that tricritical points obey the additional condition
\be
(\delta\opv\cdot\nabla)^4 f(\opv) - 3 \vv\cdot \mmat^{-1} \vv =0
\label{tricritical}
\ee
where
\[
\vv=\nabla(\delta\opv\cdot\nabla)^2 f(\opv).
\]
Even though the matrix $\mmat$ has a zero eigenvalue
($\mmat\delta\opv=\zv$ from\eq{spinodal}), the inverse of $\mmat$
in\eq{tricritical} is well defined: From\eq{critical}, we have
$\delta\opv\cdot\vv=0$, so that $\vv$ is orthogonal to the
corresponding eigenspace. Note that while the first term on the
l.h.s.\ of\eq{tricritical} is what one might have expected naively,
the second term cannot be neglected: It accounts for the fact that the
curve $\opv(s)$ passing through the three (infinitesimally separated)
stable phases is generally curved rather than straight.

We finally note that\eq{tricritical} is derived from\eq{deriv_k} for
$k=3$. In principle, the equation for $k=4$ gives an additional
condition that tricritical points must obey. Because of symmetries
present in our problem, however, this condition is always satisfied in
the cases we consider, and so we do not give its explicit form.


\end{document}